\documentclass[aps,reprint,pre]{revtex4-1}

\usepackage{graphicx}
\usepackage{color}
\usepackage{soul}

\usepackage[usenames,dvipsnames]{xcolor}
\usepackage{epstopdf}
\usepackage{dcolumn}
\usepackage{bm}

\usepackage{nomencl}
\makenomenclature

\begin{document}

\preprint{APS/123-QED}
\title{Acoustically-induced slip in sheared granular layers: application to dynamic earthquake triggering}


 \author{Behrooz Ferdowsi}
 \email{behrooz@sas.upenn.edu}
 \affiliation{%
 Swiss Federal Institute of Technology Z\"{u}rich (ETHZ) - Department of Civil, Environmental and Geomatic Engineering, Z\"{u}rich, Switzerland\\ }
 \affiliation{%
Swiss Federal Laboratories for Materials Science and Technology (Empa), ETH Domain - \"{U}berlandstrasse 129, CH-8600, D\"{u}bendorf (Z\"{u}rich), Switzerland\\}%
 \affiliation{%
Present address: Earth and Environmental Science, University of Pennsylvania, Philadelphia, PA 19104, USA
}%
\author{Michele Griffa}%
\affiliation{%
 Swiss Federal Laboratories for Materials Science and Technology (Empa), ETH Domain - \"{U}berlandstrasse 129, CH-8600, D\"{u}bendorf (Z\"{u}rich), Switzerland
}%
\author{Robert A. Guyer}
\affiliation{
Solid Earth Geophysics Group, Los Alamos National Laboratory - MS D443, NM 87545, Los Alamos, USA\\
}%
\affiliation{
Department of Physics, University of Nevada, Reno (NV), USA
}%
\author{Paul A. Johnson}
\affiliation{%
Solid Earth Geophysics Group, Los Alamos National Laboratory - MS D443, NM 87545, Los Alamos, USA
}%

\author{Chris Marone}
\affiliation{%
Department of Geosciences, Pennsylvania State University - PA 16802, University Park, USA\\
}%
\affiliation{%
G3 Centre and Energy Institute, Pennsylvania State University - PA 16802, University Park, USA
}%

\author{Jan Carmeliet}
\affiliation{%
Chair of Building Physics, Swiss Federal Institute of Technology Z\"{u}rich (ETHZ) - Wolfgang-Pauli-Strasse 15, CH-8093, Z\"{u}rich, Switzerland\\
}%
\affiliation{%
Swiss Federal Laboratories for Materials Science and Technology (Empa), ETH Domain - \"{U}berlandstrasse 129, CH-8600, D\"{u}bendorf (Z\"{u}rich), Switzerland
}%

\date{\today}

\begin{abstract}
A fundamental mystery in earthquake physics is ``how can an earthquake be triggered by distant seismic sources?''  Here, we use discrete element method simulations of a granular layer, during stick-slip, that is subject to transient vibrational excitation to gain further insight into the physics of dynamic earthquake triggering. Using Coulomb friction law for grains interaction, we observe delayed triggering of slip in the granular gouge. We find that at a critical vibrational amplitude (strain) there is an abrupt transition from negligible time-advanced slip (clock advance) to full clock advance, {\it i.e.}, transient vibration and triggered slip are simultaneous. The critical strain is order of $10^{-6}$, similar to observations in the laboratory and in Earth. The transition is related to frictional weakening of the granular layer due to a dramatic decrease in coordination number and the weakening of the contact force network. Associated with this frictional weakening is a pronounced decrease in the elastic modulus of the layer. The study has important implications for mechanisms of triggered earthquakes and induced seismic events and points out the underlying processes in response of the fault gouge to dynamic transient stresses. 

\end{abstract}

\maketitle


\section{Introduction}

Dynamic triggering of earthquakes by seismic waves is a robustly observed phenomenon that is well-documented for over 30 major earthquakes worldwide~\citep{brodsky2014uses} and many more smaller earthquakes~\citep{Gomberg2001,Gomberg2004,Gomberg2005,gomberg2014crustal}. Recent observations based on new, more sensitive instrumentation show that a majority of earthquakes may be dynamically triggered~\citep{vanderElst2010,Marsan2008}. Laboratory-scale experiments and seismological observations indicate that a key role in dynamic earthquake triggering may be played by granular materials, termed ``fault gouge", accumulated at the core of a geologic fault~\citep{Johnson2008,Johnson2005}. The observations at the laboratory and field scales strongly suggest that the nonlinear dynamical response of the gouge material significantly contributes to triggering, although details remain unquantified. Direct access to the earthquake fault gouge without changing its microstructure and loading history is not possible. However, we here aim to characterize the granular physics of triggering on laboratory scales using physical experiments and numerical simulations.

Granular layers exhibit stick-slip dynamics when they are subjected to shearing, at sufficiently high confining pressures and low shearing velocities~\citep{Aharonov2004,Aharonov1999,Mair1999,lieou2015stick,ciamarra2009granular,Griffa2011}. The stick-slip instabilities have been associated with a non-monotonic shear stress {\it vs} shear strain response of granular materials that have frictional constituents or frictional dissipation~\citep{lieou2015stick}. The stick-slip dynamics are analogous to the seismic cycle in earthquake fault systems. Fault systems accumulate strain energy during the interseismic period of the seismic cycle, just as a sheared granular layer does during the stick phase of the stick-slip cycle~\citep{Brace1966, Johnson1973}. Laboratory scale observations confirm that mechanical vibrations with adequate amplitudes can change the mechanical and frictional properties of the granular layer, changing its macro-scale response. This includes a transition from a solid-like behavior to a transient, fluid-like one~\citep{Luding1994,Savage2008,Janda2009,Capozza2009,Melhus2009,Jia2011,Xia2013}. The behavior of granular materials under different loading conditions and to different perturbations is controlled by their evolving internal structure including the contact force networks, particle rearrangements and force distribution between the particles inside the granular layer~\citep{cates1998jamming,majmudar2005contact,hunt2010force}.~\citet{Jia2011} identified two regimes of fast nonlinear dynamics versus the input amplitude. In the first regime, the interaction between sound waves and the granular medium is reversible: during the wave excitation the modulus can change. However, neither velocity nor sample density are changed after the wave passage and the force network remains nearly unchanged. In the second regime, beyond a certain amplitude threshold depending on the applied load, the sound-matter interaction becomes irreversible. In addition, the wave velocity and corresponding elastic modulus remain weakened after the wave transient, and permanent deformation is observed corresponding to an accompanying compaction. This finding highlights the relationship between the macroscopic elastic weakening and the local change of the contact network, induced by strong sound vibration in the absence of visible grain motion~\citep{Jia2011}.

~\citet{Johnson2008} observed both instantaneous and delayed triggered (cascading) slip in the lab, when vibration amplitudes corresponding to strains $>\sim10^{-6}$ are applied at shear stress levels of $\approx 95 \%$ of the failure value. Other studies also demonstrated the existence of threshold values of strain amplitude for dynamic earthquake triggering. The existence of a unique strain threshold value is an important open question~\citep{vanderElst2010}, however there is increasing evidence that in many cases dynamic earthquake triggering may be governed by such a threshold mechanism~\citep{Pollitz2012,Gomberg2005,Gomberg2001}. ~\citet{Johnson2008} also observed other features in common with earth faults including disruption in the earthquake recurrence interval (the time interval between earthquakes) in response to dynamic perturbations, as well as triggering-induced changes in the gouge material modulus. We have previously investigated the deformation characteristics for dynamically triggered slip and the influences of vibration amplitude on triggering using two dimensional discrete element simulations of a granular fault gouge\citep{Griffa2012, Griffa2013}. In a follow-up study, we developed a three dimensional discrete element model of the granular fault gouge that showed similar dynamics to earthquake stick-slip cycles and aseismic creep prior to slip events\citep{ferdowsi2013b,Ferdowsi2014}. In that system, we have characterized the short and long term influences of triggering with regard to stick-slip size distribution and recurrence intervals\citep{Ferdowsi2014three,ferdowsi2014discrete}. Here, we report results of three dimensional discrete element method simulations of granular gouge layers subjected to boundary vibrations to understand the grain-scale mechanics of dynamic earthquake triggering and the existence of a triggering threshold under certain confining pressures. We also explore the nature of the transition to significant clock-advanced triggered slip events at different triggering vibration amplitudes and frequencies. This modeling work is set to complement experimental observations obtained from double-direct shear experiments by~\citet{Johnson2008}. 

\section{Model setup}
\begin{figure}
\noindent\includegraphics[width=10pc]{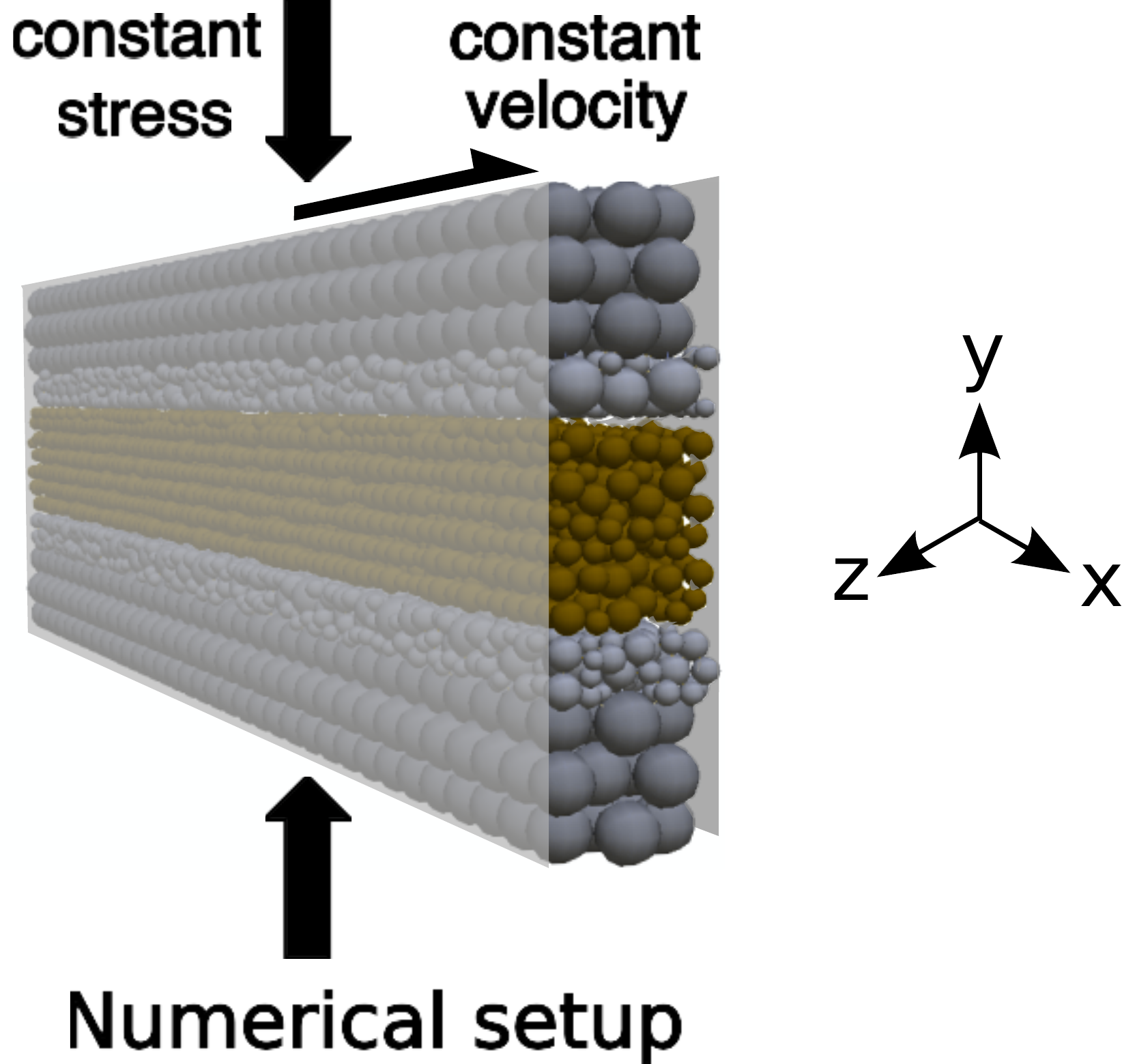}
\caption{3D DEM model comprised of the driving block (top), a granular gouge layer (center) and a substrate block (bottom)}
\label{fig:Fig1}
\end{figure}

Figure~\ref{fig:Fig1} illustrates the simulated granular gouge layer. The model consists of three layers of particles: a driving block at the top, a granular gouge layer and a substrate block at the bottom. The driving and substrate blocks are used to confine the granular gouge by applying a constant normal force in the $Y$-direction. The top driving block moves at constant velocity in the positive $X$-direction and applies a shear force to the granular gouge layer. Each variable/parameter in our 3D DEM model is expressed in terms of the following basic dimensional units: 
$L_0 = 150\ \mu m$, $t_0 = 1\ s$ and $M_0 = 1\ kg$, for length, time and mass, respectively. $L_0$ represents the largest particle radius within the overall DEM model.  We run sheared granular layer simulations at a confining pressure of $\sigma_n=6000 \frac{M_0}{L_0{t_0}^{2}}$ (40 MPa) and shearing velocity of $V_{X,0} = 0.004 \frac{L_0}{t_0}$ (0.6 $\frac{\mu m}{s}$) to achieve stick-slip dynamics. We chose parameter values to match laboratory experiments, rather than tectonic fault zones, although the two overlap in many ways. Further details about the model are provided in the supplementary materials. 

The simulations that are not subjected to vibration are called ``reference'' runs, while those with vibration are called ``perturbed'' runs. In the perturbed (also called ``triggered") simulation runs, an additional boundary condition consists of imposing a cyclic displacement in the $Y$ direction for the bottom particles of the substrate for the duration of about $\Delta t = 0.1$ ${t_0}$. The characteristic of the vibration signal is shown in figure ~\ref{fig:Fig2}-a (gray zone). The temporal displacement of the boundary vibration is described in the supplementary materials. Perturbations with longer duration have a bigger influence than those with shorter duration. The influence of vibration duration is provided in the supplementary materials. Vibration normal to the boundary (displacement in $Y$ direction) is found to be more effective in triggering slip than horizontal vibration (in $X$ and $Z$ directions) primalry due to better transmission of normal vibration to and through the cohesionless granular gouge layer compared with the horizontal vibration. The comparison of horizontal and vertical vibration influences is included in the supplementary materials. For the DEM model, we use a range of vibration amplitudes including A =$\{1, 10, 20, 30, 40, 50, 60, 70, 80, 90, 100\}\times10^{-7}L_0$. We can estimate the strain induced by the vibration as, $\epsilon=\frac{A}{\lambda_{vib}}$, where $\epsilon$ is the induced strain, $A$ is the vibration amplitude in [$L_0$] unit, and $\lambda_{vib}$ is the vibration wavelength. The wavelength is $\lambda=\frac{\nu}{f_{vib}}$, where $\nu$ is the sound speed and $f_{vib}$ is the vibration frequency\footnote{Here, the density of the granular layer is, $\rho\approx1.8339\cdot10^{11}$ [$\frac{kg}{m^3}$]. The density of the granular layer, $\rho$, is calculated as the ratio of the mass of the total number of particles to the volume of the granular gouge layer at its sheared confined state. The density of a single particle in our DEM model, $\rho_s$, is chosen as $2.9\times10^{11}[\frac{kg}{m^3}]$. The choice of $\rho_s$ is based on a density scaling scheme \cite{cundall1982,thornton2000,osullivan2004} that is frequently used in DEM modeling studies to increase the simulation time step ($\Delta t \propto \sqrt{\rho_s}$) and to make the simulations computationally feasible. At the same time, we checked that the Inertia number \cite{midia2004dense,Agnolin2007,sheng2004} of the DEM runs is always below $10^{-6}$ to ensure that the behavior remains in the quasi-static regime and the density scaling effects would be negligible. The sound speed is approximately $\nu=\sqrt{\frac{K}{\rho}}$. The bulk modulus, $K$, is measured at different triggering times and is ranging from 10-40 GPa.}.

The reference 3D DEM model behavior are described in an earlier work~\cite{ferdowsi2013b}. The packing fraction in the simulation is $\sim$ 0.58 at the beginning of stick phases. The packing fraction gradually decreases while the granular layer dilates during the stick-phase to $\sim$ 0.56.  The stick-slip behavior is monitored by its friction coefficient time series. The friction coefficient, $\mu$, is defined as the ratio of shear stress developed at the boundary layers to the imposed normal stress. 

\section{Results}

n our DEM model as well as the experimental setup studied by~\citet{Johnson2008}, vibration amplitudes that induce a strain value of order $\sim10^{-6}$ cause a time-advanced slip (clock advance)~\citep{Ferdowsi2014three}. In both setups, the vibration causes an immediate weakening (reduction of shear strength) of the granular layer

\begin{figure*}
\noindent\includegraphics[width=36pc]{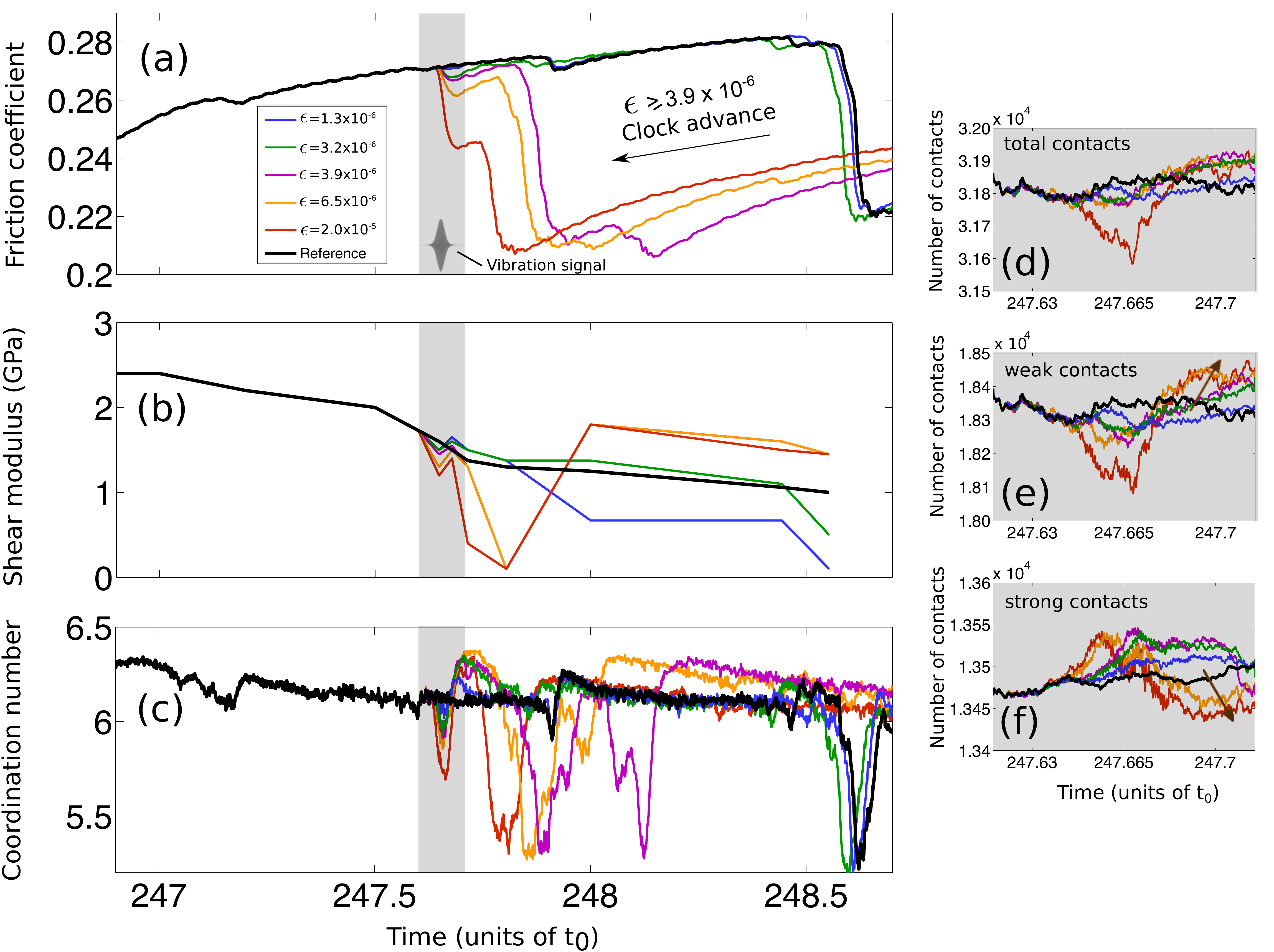}
\caption{(a) Friction coefficient signal, (b) shear elastic modulus measurements, (c) coordination number variations for reference (black) and perturbed simulations with a range of vibration strains. The vibration interval is shaded in all panels. subfigures (d), (e), and (f) show the variation of total number of contacts, number of weak contacts and number of strong contacts, respectively, during the full vibration interval.}
\label{fig:Fig2}
\end{figure*}

Figure~\ref{fig:Fig2}-a shows the behavior for selected vibration amplitudes in the DEM model. Here, vibration amplitude larger than $\sim6\times10^{-6}L_0$ (corresponding to induced strain of $\sim3.9\times10^{-6}$) causes a sharp clock advance.  ~\citet{Johnson2008} report a highly perturbed stick-slip recurrence interval due to acoustic excitations, compared to the reference case. Similarly in the simulation, vibration induced clock advance means that the recurrence time for the next event will be longer. In addition, \citet{Johnson2012} report a similar vibration strain for inducing slow slip event in sheared granular fault gouge. We measure shear elastic modulus of the granular layer to monitor the evolution of elastic properties corresponding to the observed clock advance. The shear modulus of the granular layer is determined by applying a small shear strain cycle to the system both in the reference and in the perturbed simulations. For this purpose, the shearing of the granular layer is stopped and the system is given 10000 numerical time steps to relax from the shearing influences before performing the shear modulus measurements. We ensured that the average normal and tangential contact forces remain constant following the relaxation process. In addition, there have been no observation of any failure event in the shear and normal stress signals and therefore no significant change in the contact network of the granular layer during this process. We then applied a cyclic shear strain to the top of the granular layer. We investigated the influence of different cyclic shear strain amplitudes for the modulus measurements. The results indicate that when the maximum applied shear strain $\gamma_{max} < 9\times10^{-6}$, the shear modulus measurements are similar to each other. When the applied shear strain is further increased, the behavior becomes nonlinear. This observation suggests that shear strain values $\gamma_{max} \ge 1\times10^{-5}$ induce considerable particle contact rearrangement in the granular gouge layer. We therefore performed the shear modulus measurements with a load cycle of maximum strain amplitude $\gamma_{max} \approx 8.6\times10^{-6}$. Furthermore, the amplitude and duration of the protocol is selected such that it does not change contacts population during its application. A similar loading convention for modulus measurements has been suggested before by~\citet{Nguyen2011}. The shear modulus is determined by fitting a line to the initial unloading part of the stress-strain curve. The procedure is explained in the supplementary materials.

Figure~\ref{fig:Fig2}-b shows the shear modulus measurements for the reference and the perturbed simulations over the range of vibration amplitudes. During vibration, we observe a decrease in the shear elastic modulus of the granular layer, indicating the mobilization of particles during the vibration interval. As the vibration amplitude increases, the shear modulus further decreases during the vibration interval. If the vibration amplitude is small (for this example, $\epsilon < 3.9\times10^{-6}$), the shear modulus recovers after the vibration terminates. However, if $\epsilon \ge 3.9\times10^{-6}$, the shear modulus does not recover at the end of the vibration interval, but decreases further as the shearing continues beyond the vibration interval. This behavior leads to the clock advance of the large expected slip event. This observation supports the hypothesis of~\citet{Johnson2005} that, upon the application of sufficiently large vibration strains, the shear modulus of the granular medium decreases and the slip event occurs with a clock advance compared to its reference time. 

The variations of the coordination number ($CN$), the number of contacts per particle, for reference and selected perturbed runs are shown in Figure~\ref{fig:Fig2}-c. The $CN$ decreases slightly in the reference run as we approach the slip onset. During slip, the $CN$ drops significantly. In the case of perturbed runs, the $CN$ decreases significantly during the vibration interval. It recovers after vibration terminates to the reference level if the vibration strain is $\epsilon < 3.9\times10^{-6}$. However, for vibration strains $\epsilon \ge 3.9\times10^{-6}$, the $CN$ initially recovers during the gradual removal of vibration, but it begins to decrease again leading to the clock advance of the slip event. Here, we note that the decreasing shear modulus of the granular gouge layer at the time of  (triggered) slip as well as low values of $CN$ are reminiscent of the response of an unjammed granular layer. However, we cannot determine whether the system is jammed or shear-jammed \citep{Bi2011} during stick phases because of the continuous slow rearrangement of the granular gouge layer that take place in the stick phase. Understanding the nature of the transition of the frictional granular layer from sticking to slipping will require further studies.

\section{Discussion}

The decrease of the $CN$ despite the initial recovery is in agreement with the change in elastic shear modulus of the granular gouge layer. We now investigate the evolution of the contact network of the granular gouge layer. The change in the total number of contacts during the vibration interval is shown in figure~\ref{fig:Fig2}-d and indicates a decrease of the total number of contacts as the vibration intensifies. The grain contacts show a slow recovery toward the end of the vibration interval for all vibrational strains.  

We now investigate the variation of the number of weak and strong contacts in the granular gouge layer. Weak contacts are those contacts that are carrying normal forces smaller than the average normal contact force of the granular gouge layer. Strong contacts are carrying normal forces larger or equal to the average normal contact force of the granular gouge layer. The contact force distribution in granular systems is approximately an exponential distribution \citep{majmudar2005contact}. Strong and weak contacts form two subnetworks of the granular contact network with complementary mechanical properties ~\citep{radjai1996force,radjai1998bimodal}. Strong contacts form the majority of the force chains or force bearing structure of the granular medium whereas weak contacts primarily support the strong contacts, distribute the force over the whole granular contact network and help mobilizing the overall granular medium. The number of weak and strong contacts during the vibration interval are shown in figure~\ref{fig:Fig2}-e and~\ref{fig:Fig2}-f, respectively. Figure~\ref{fig:Fig2}-e shows that for vibration strains $\epsilon \ge 3.9\times10^{-6}$, the weak contact number decreases dramatically at peak of vibration. The number of strong contacts on the contrary increases at peak vibration for $\epsilon \ge 3.9\times10^{-6}$ as can be observed in figure~\ref{fig:Fig2}-f. The weak contacts regenerate and recover as the vibration is removed, while the number of strong contacts decreases for vibration strains $\epsilon \ge 3.9\times10^{-6}$. These observations suggest that the strong contact network is strengthened when the vibration strain increases initially. This is more or less expected, because we are applying an excess confining pressure due to the vibration. Indeed, we observe some small scale fluctuations in the strong contact number synchronized with the applied vibration. Following the peak in the vibration strain, we observe a weakening of the strong contact network. After peak vibration, the weak contact recovers and strengthens above the reference level. This fact could explain the occurrence of delayed dynamic triggering: the strong contact network population is permanently depleted of members by the vibration application. It implies that the system is more susceptible to grain-scale rearrangements in correspondence of the continuously applied shear load, since the granular medium is in a critical state close to failure. This result indicates that the large vibration strain weakens the strong contact network of the granular layer and by this causes a clock advance of the expected slip event. 

\begin{figure}
\noindent\includegraphics[width=20pc]{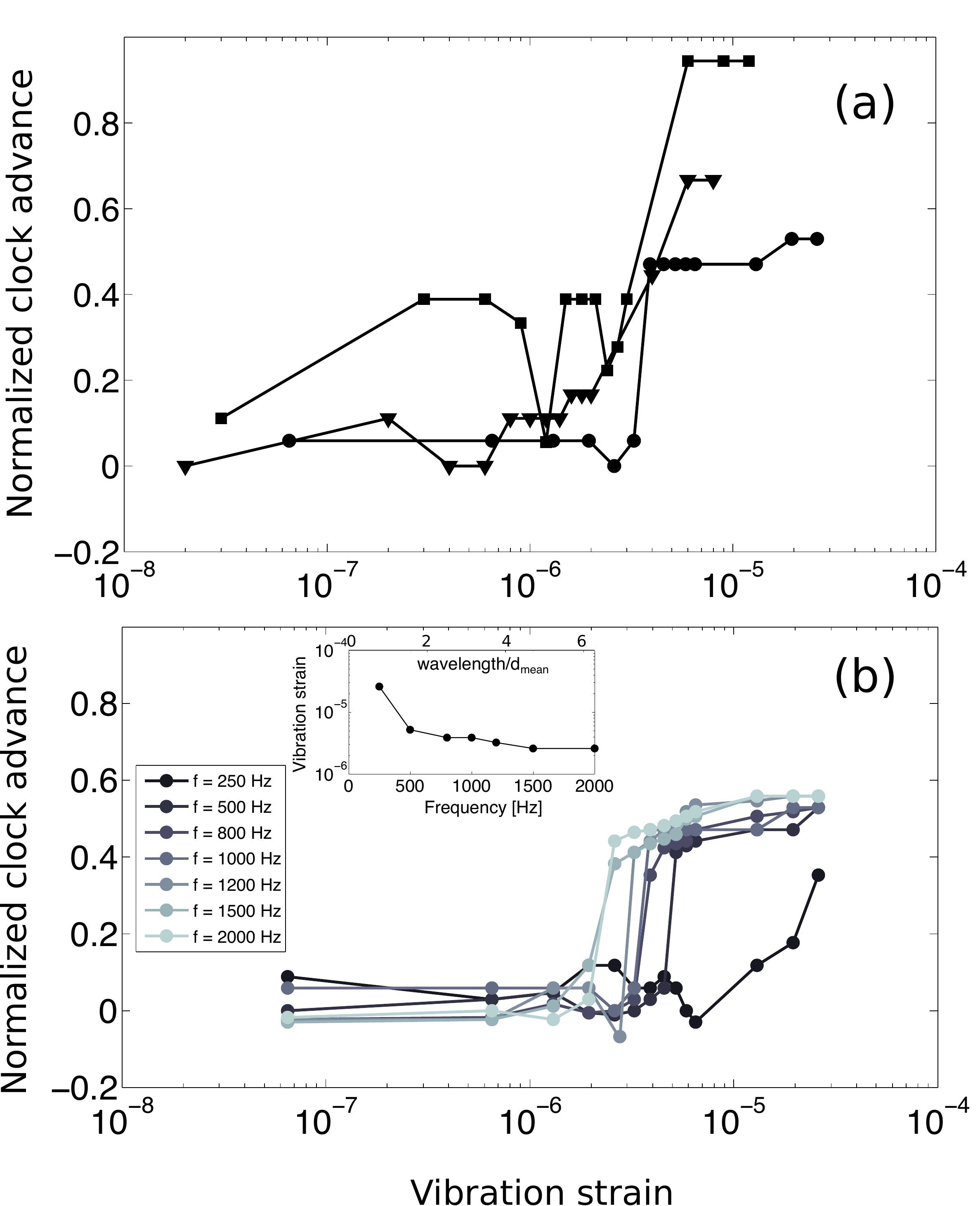}
\caption{(a) Normalized clock advance time of triggered slip, defined as the difference between the reference and triggered slip times divided by the recurrence interval of the reference slip event, for three different reference slip events shown with symbols (circle), (triangle) and (square) at a range of vibration strains. (b) Normalized clock advance time of triggered slip for a range of vibration strains at different triggering frequencies. Inset: minimum vibration strain required for causing a significant clock advance at different vibration frequencies and ratios of vibration wavelength to mean particle diameter in the gouge layer corresponding to these frequencies (top axis of the inset).}
\label{fig:Fig1c}
\end{figure}

We further explore the nature of the transition to significant clock advance for a range of vibration strains and frequencies. Figure~\ref{fig:Fig1c}-a shows the normalized clock advance time, defined as the difference between the reference and triggered slip times divided by the recurrence interval of the reference slip event, for a range of vibration strains applied in $\sim70\%$ of the stick-phase of three different stick-slip cycles. This three stick-slip cycles belong to the same numerical setup, but is chosen from different temporal points in the stick-slip time-series of the system. The clock advance {\it vs.} vibration strain plot shows characteristics of a first order transition that takes place at vibration strain $\epsilon \sim 10^{-6}$ for all studied events, however the amount of clock advance can be different for different slip events. Due to the discontinuous nature of the granular materials, the fault gouge layer is sensitive to the vibration frequency, {\it i.e.} the frequency at which a wave packet explores the medium, introducing a competition between rearrangement length scale and vibration wavelength. We know from seismotectonics that an earthquake fault rupture generates seismic waves with a spectrum of frequencies, which influences the granular gouge of nearby or distant fault zones. It is thus of relevance to understand the consequences of triggering at different vibration frequencies. This is shown in Fig.~\ref{fig:Fig1c}-b where we perturb the granular gouge layer at a range of frequencies from 250 to 2000 Hz. The transition is a first order type process for different vibration frequencies. The vibration strain at which the transition occurs decreases with increasing the vibration frequency to about $f_{vib}=1500$ Hz. The inset of Fig.~\ref{fig:Fig1c}-b shows the minimum vibration strain required for a significant clock advance versus the frequency of the applied vibration. This figure shows that by increasing the frequency of vibration, the minimum vibration strain required for observing a clock advance of the expected large slip event gradually decreases until we reach a vibration frequency of $\sim$800 Hz. From this point on by further increasing the vibration frequency the minimum amplitude for clock advance only decreases slightly and stays at the vibration amplitude of $\epsilon \ge 4\times10^{-6}$ for the vibration frequencies of 1500 and 2000 Hz.

This trend can be explained by calculating the vibration wavelength. For example, for the vibration frequency $f_{vib} = 1000$ Hz, we have  $\lambda_{vib} = \frac{\nu}{f_{vib}} = \frac{0.2}{1000} = 2\times10^{-4} m$, where $\lambda_{vib}$ and $\nu$ are the vibration wavelength and speed of sound in the granular layer, respectively. The grain diameter in the granular gouge layer is in the range of $[1.05; 1.5]\times 10^{-4} m$. Thus we find that the vibration wavelength for the vibration frequency  $f_{vib}=1000$ Hz is very close to the average size of the granular gouge particles. Further increase of the vibration frequency decreases the vibration wavelength. The wavelength at frequencies in the range of 800 to 1000 Hz can thus explore the granular media at the grain scale and perturb the contact network of the medium more effectively. However, decreasing the vibration frequency increases the vibration wavelength. As a result, the perturbing vibration at frequencies lower than 500 Hz, are at length scales larger than the grain size and cannot significantly perturb the grain arrangements and the contact network of the granular layer. A similar behavior for the triggering frequency has been observed by \citet{savage2007effects} in laboratory experiments with sheared and dynamically perturbed granular layers, where they notice that upon increasing the triggering force frequency, the friction coefficient at dynamic (perturbed) failure starts to decrease compared to the friction coefficient at reference (unperturbed) failure until a critical frequency is reached. By further increase of the frequency beyond that critical value, the ratio of the friction at dynamic failure to the friction at the reference failure saturates~\citep{savage2007effects}. The vibration frequency $f_{vib}=1000$ Hz used in this study is in a range that can effectively perturb the granular layer at the grain scale and change its contact network significantly with short time exposures.

\section{Summary}

We have investigated the influence of boundary vibration on stick-slip dynamics of sheared granular layers using discrete element simulations. The numerical work is meant to complement the experimental observation by~\citet{Johnson2008}. In both discrete element simulations and experiments, a range of vibration strains is used for slip triggering and it is shown that above a critical amplitude of about $\sim10^{-6}$, vibration causes a significant clock advance of large slip events. Numerical simulation shows that vibration initially influences the network of weak contacts in the granular gouge layer. The weak contacts can be regenerated after removal of vibration, whereas strong contacts remain weakened compared to their pre-vibration state. We link the observed clock advance for $\epsilon \geq \sim10^{-6}$ to a major decrease of the coordination number as well as weakening of the strong contact network of the granular gouge layer. We further found that clock advance of the triggered slip event is a first order phase transition process in correspondence with increasing vibration strain amplitude. We would like to emphasize that in the Earth, the exposure time to dynamic stresses, sound velocity in the fault gouge and fluid pressure are among the factors that determine the effectiveness of triggering. The triggering threshold, if exists in Earth and field, may depend on the type of interaction between grains, for example on the existence of cohesion, humidity and fluid pressure (\citet{scuderi2015poromechanics}), as well as on the physical and chemical properties of contact asperities and gouge roughness. These are among the areas that need further research and investigation. 

\begin{acknowledgments}
We thank D. Weatherley and S. Abe for support with the implementation of our DEM model in the \href{https://launchpad.net/esys-particle}{ESyS-Particle} code and D. Passerone and C. Pignedoli for the help related with the use of the \href{http://www.hpc-ch.org/empa/}{High Performance Computing cluster} at Empa. Our work has been supported by the Swiss National Science Foundation (projects No. 206021-128754 and No. 200021-135492), by funding from the DOE Geothermal office and the LDRD Program (Institutional Support) at the Los Alamos National Laboratory, USA. 
\end{acknowledgments}

\bibliography{libraryPRE}

\end{document}